\documentclass[aps,pra,showpacs,twoside,twocolumn,longbibliography,10pt]{revtex4-2}
\usepackage[colorlinks=true, citecolor=red, urlcolor=blue ]{hyperref}
\usepackage{epsfig,newlfont,amssymb,amsfonts,amsmath,bm,subfigure,palatino,mathtools,amsthm,braket,times,soul,enumitem,color}
\usepackage[normalem]{ulem}
\usepackage[english]{babel}
\usepackage[utf8]{inputenc}
\usepackage{array}
\usepackage{xcolor}
\usepackage{graphics}
\usepackage{amsthm}
\usepackage{verbatim}
\usepackage{physics}
\usepackage{bbm}
\usepackage{wrapfig}

\usepackage{multirow}
\usepackage{hyphenat}
\usepackage{graphicx} 
\usepackage{mathrsfs}
\usepackage{bm} 
\usepackage{mleftright}
\DeclareMathAlphabet{\mathbbmsl}{U}{bbm}{m}{sl}

\definecolor{med-blue}{RGB}{25,25,112} 
\hypersetup{colorlinks, linkcolor={blue},citecolor={blue}, urlcolor={blue}}

\newcommand{\outpr}[2]{\mleft\vert{#1}\mright\rangle\mleft\langle{#2}\mright\vert}

\newcommand{\expv}[2]{\mleft\langle{#1}\mright\rangle_{#2}}

\newcommand{\proj}[1]{\outpr{#1}{#1}}

\begin{document}

\title{
Noise cancellation by superposition of channels and superactivation of quantum capacity: Experimental realization by NMR
}

\author{Deepika Bhargava}
\email{deepika.bhargava@students.iiserpune.ac.in} 
\affiliation{Department of Physics and NMR Research Center, \\Indian Institute of Science Education and Research, Pune 411008, India}

\author{Arijit Chatterjee}
\email{arijit.chatterjee@ens-lyon.fr} 
\affiliation{Department of Physics and NMR Research Center, \\Indian Institute of Science Education and Research, Pune 411008, India}

\author{Vishal Varma}
\email{vishal.varma@students.iiserpune.ac.in}
\affiliation{Department of Physics and NMR Research Center, \\Indian Institute of Science Education and Research, Pune 411008, India}

\author{T. S. Mahesh}
\email{mahesh.ts@iiserpune.ac.in}
\affiliation{Department of Physics and NMR Research Center, \\Indian Institute of Science Education and Research, Pune 411008, India}

\begin{abstract}
Noisy quantum channels degrade quantum resources such as coherence and entanglement and hence pose  challenges for realizing quantum technologies.  Coherent control of noisy channels allows us to minimize their effects on the quantum system.
Here we achieve the cancellation of two noisy quantum channels by superposing their corresponding Stinespring dilation unitaries. We first arrive at conditions under which superposition of channels results in a valid quantum channel. We then consider superposing two dephasing channels and observe their destructive interference, thereby effectively recovering the quantum coherence. On superposing two zero-capacity depolarizing channels, we show superactivation of quantum capacity. 
We experimentally realize the cancellation of two dephasing channels using a three-qubit NMR register.   Furthermore, using a five-qubit NMR register, we realize the cancellation of two depolarization channels and demonstrate superactivation of quantum capacity.

\vspace{5pt}
\end{abstract}

\maketitle
\section{Introduction}
One of the most exciting elements of quantum physics is the possibility of coherent superposition of two or more distinct states of a physical system.  Likewise, it also allows two or more quantum processes to be in superposition \cite{interference-q-channels, roy2012nmr, vydp-9qqq}. The act of superposition of two processes can cause their spatial or temporal order to be indistinguishable \cite{ICO_1}. This was originally hypothesized in the context of gravitational fields \cite{ICO_gr, ICO_gr2}, and  later adopted to quantum information through the introduction of the quantum switch \cite{qswitch_1}. The quantum switch places the processes in a temporal superposition by superposing the order in which they occur. This has already been explored for applications in a variety of tasks, such as quantum communication \cite{qswitch-comm-1, qswitch-comm-2} and metrology \cite{qswitch-metero1, qswitch-metero2, qswitch-metero3}. Several experimental realizations of  the quantum switch have also demonstrated its applications \cite{Procopio2015, qswitch2_exp, doi:10.1126/sciadv.1602589}. 
Likewise, the spatial superposition of channels is achieved by coherent control of trajectories that lets a quantum system evolve under multiple channels simultaneously \cite{interference-q-channels, sup-traj1}. Spatial superposition of unitary channels promises applications in quantum communication, quantum information, quantum emulations \cite{roy2012nmr, vydp-9qqq}, and quantum thermodynamic tasks \cite{sup-traj-thermo}. On the other hand, two or more nonunitary channels can also interfere, resulting in their cancellation.  For instance, in the phenomenon of cross-correlation in NMR, natural relaxation processes interfere and cancel, which was exploited in high-resolution protein NMR spectroscopy \cite{cross-corr-nmr1, cross-corr-nmr3, salzmann1998trosy}. A quantum system can also evolve under a superposition of nonunitary channels.  There have been many attempts to compare the quantum switch with the superposition of trajectories in different settings \cite{sup-traj-vs-switch1, sup-traj-vs-switch2, sup-traj-vs-switch3}. 

\begin{figure}
    \centering
    \includegraphics[width=0.95\columnwidth]{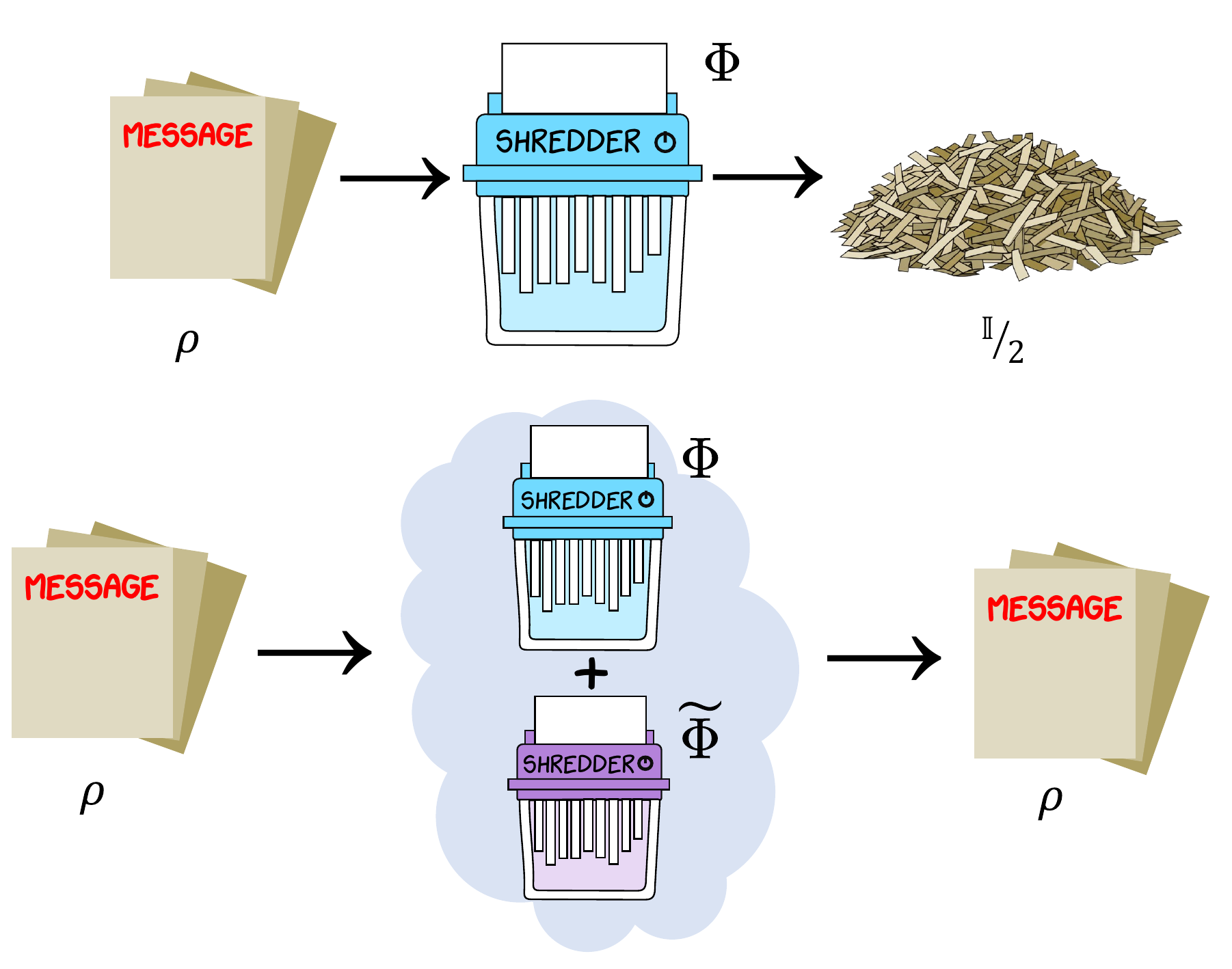}\\
    \caption{Illustrating the cancellation of two channels when superposed.}
    \label{abstract-img}
\end{figure}

In this work, we first describe a framework to engineer an interference of two quantum channels using a common ancillary register and arrive at conditions for realizing a valid resultant channel.
For single qubits, the general Pauli channels  constitute the fundamental model for simulating quantum noise \cite{gen-pauli1, gen-pauli2}.  These channels are used to improve error-correcting techniques for fault-tolerant quantum computing \cite{pauli-channel-error}. We show that the controlled superposition of two such channels forms another Pauli channel whose strength depends on the superposition parameter. We then study two specific examples of general Pauli channels, namely dephasing and depolarizing channels, and discuss their superpositions.
Using a three-qubit NMR register, we experimentally realize the interference of two dephasing channels and demonstrate the recovery of quantum coherence.
We then focus on another subclass of Pauli channels, namely entanglement breaking (EB) channels \cite{EBC, qubit-EBC}. They are extensively used in quantum technology in the context of quantum memories \cite{EBC-quantum-memory1, EBC-quantum-memory2, EBC-quantum-memory3} and quantum communication to facilitate the transmission of information \cite{EBC-comm1, EBC-comm2}. Previous models involving the quantum switch have shown how superposing two EB channels can result in superactivation of quantum capacity, thereby boosting communication \cite{EBC-comm1, EBC-comm2}. However, it requires more resources in the form of ancillary qubits and quantum control. Here, we discuss superactivation as well as perfect activation of quantum capacity by superposing two EB channels with fewer resources.  Using a 5-qubit NMR register, we experimentally realize a non-EB channel with positive quantum capacity from the superposition of two zero-capacity EB channels, thus demonstrating superactivation (see Fig. \ref{abstract-img}).

The article is structured as follows. In Sec. \ref{sup-channel}, we describe the framework for superposition of channels, analytically arrive at the conditions to ensure a valid superposed channel, and discuss the superposition of general Pauli channels. In Sec. \ref{dephasing-super-sect} we describe the specific example of superposing two dephasing  channels and its experimental demonstration.  In Sec. \ref{depol-super-sect} we discuss superposing two EB channels and report experimental realization of superactivation.  Finally, we conclude in Sec. \ref{conclusions}

\section{Superposition of Channels}
\label{sup-channel}
The quantum channel $\Phi$ acting on a system initialized in state $\rho_S$ can be modelled as a joint unitary evolution $U$ of the system and environment, followed by tracing out the environment \cite{Nielsen_Chuang_2010}. 
Even though for simplicity our construction of the Stinespring dilation assumes the ancillary qubits to be initialized in a pure state, the framework extends mutatis mutandis to mixed states via a simple state purification.
To realize a channel with $m$ Kraus operators, we use $n = \lceil \log_2 m \rceil$ or more ancillary qubits to act as the $2^n$-dimensional environment.  
Here, the extra $2^n - m$ dimension of the environment corresponds to null Kraus operators.  
Thus, the channel is defined by,
\begin{equation}
    \Phi(\rho_S) = \Tr_A(U(\rho_A \otimes \rho_S)U^\dagger),
\end{equation}
wherein one normally takes $\rho_A = \ketbra{0_n}{0_n}$, with $\ket{0_n} = \ket{0^{\otimes n}}$, as the initial state of the ancilla.  Such a model, called the Stinespring dilation  of the channel $\Phi$ \cite{Stinespring},  
forms a completely positive and trace-preserving (CPTP) map having an operator-sum representation,
\begin{equation}
    \Phi(\rho_S) = \sum_{i=0}^{m-1} K_i\rho_S K_i^\dagger,
\end{equation} 
with Kraus operators $K_i = \bra{e_k}U\ket{0_n}_A$ and ancillary basis vectors $\{e_k\}$.
 
\begin{figure}
    \centering
    \includegraphics[width=0.9\columnwidth]{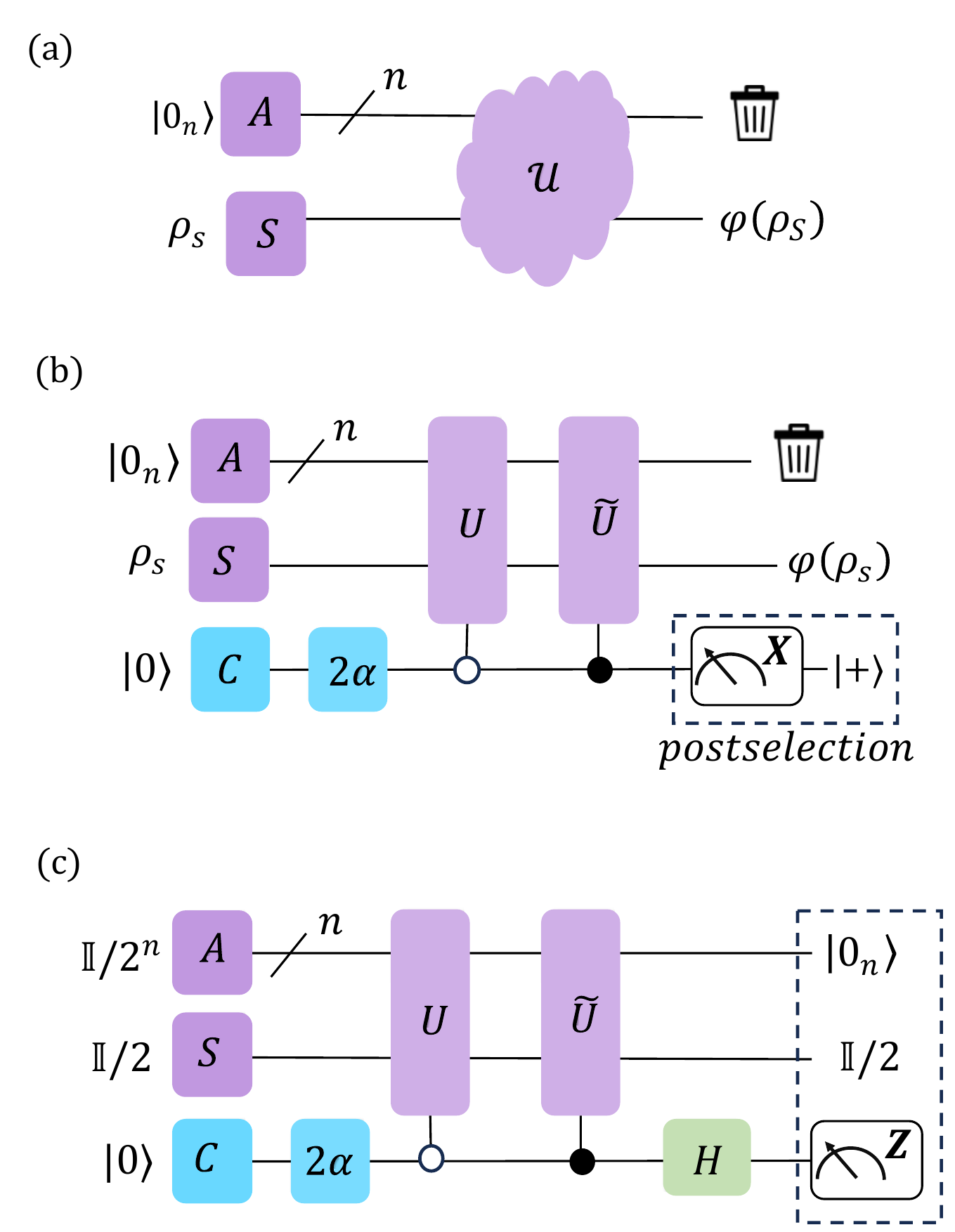}
    \caption{(a) Realizing superposition of channels on system $S$ via a superposed unitary acting on a larger space, followed by partial trace of ancillary register $A$. (b) Full quantum circuit for realizing superposition of channels via two Steinspring dilation unitaries $U$ and $\widetilde{U}$ controlled by qubit $C$, followed by postselection and partial trace.
    (c) Quantum circuit for finding the normalization factor $\gamma$. 
    It involves a conditional measurement of diagonal elements as shown.}
    \label{fig:circuits}
\end{figure}
Like a qubit prepared in the state $\ket{\psi} = 
\cos{\alpha}~\ket{0} + \sin{\alpha}~\ket{1}$
is regarded as a superposition of 
basis states $\set{\ket{0},\ket{1}}$, Arijit et al. \cite{vydp-9qqq} have defined the superposition of unitaries $U$ and $\widetilde{U}$ as 
\begin{align}
\mathcal{U} = (\sin{\alpha}~U + \cos{\alpha}~\widetilde{U})/\mathcal{N},~
\end{align}
where $\mathcal{N}$ is the normalization factor such that 
$\mathcal{U}\mathcal{U}^\dagger = \mathbbm{1}$.

In this work, we extend the above framework to the superposition of quantum channels.  
One way to extend is to implement a superposed Stinespring dilation unitary acting on an extended space of system and ancilla, followed by tracing out ancilla as illustrated in Fig. \ref{fig:circuits} (a).
To realize the superposed channel $\varphi$ of channels $\Phi$ and $\widetilde\Phi$ with Kraus operators $\set{K_i}_{i=0}^{m-1}$ and  $\set{\widetilde{K}_i}_{i=0}^{m-1}$ modelled by Stinespring unitaries $U$ and $\widetilde{U}$, a control qubit $C$ is initialized in the state 
$\ket{\alpha} = \cos{\alpha}\ket{0} + \sin{\alpha}\ket{1}$ and the operation \begin{equation}
\mathbbm{U} = U_{AS}\otimes \ketbra{0}{0}_C + \widetilde{U}_{AS}\otimes\ketbra{1}{1}_C    
\end{equation}
is applied on the joint space of the ancilla $A$, system $S$, and control $C$. 
To construct the unitaries $U$ and $\widetilde{U}$, we stack the Kraus operators in a column to create an isometry and extend it into a unitary by adding orthonormal vectors \cite{Nielsen_Chuang_2010}. Finally, $\ket{+}$ outcome of $C$ is postselected after measuring it in the $\set{\ket{\pm}}$ eigenbasis of Pauli-X operator,  and the ancilla is traced out. The corresponding quantum circuit is shown in Fig. \ref{fig:circuits} (b).

\subsection{Conditions for a valid superposed channel}
\label{sec-validsup}
The final state of the joint register after the evolution under the controlled unitaries shown in Fig. \ref{fig:circuits} (b) is,
\begin{align}
    \label{final-system-state-eqn}
    \rho^f 
    &= \mathbb{U}\!\left(
         \rho_{AS} \otimes \ketbra{\alpha}{\alpha}_C
      \right)
      \mathbb{U}^\dagger \nonumber \\[4pt]
    &= a \otimes \ketbra{0}{0}_C
      + b \otimes \ketbra{0}{1}_C
      + b^\dagger \otimes \ketbra{1}{0}_C
      + c \otimes \ketbra{1}{1}_C \nonumber \\[4pt]
    &= \tau \otimes \ketbra{+}{+}_C + \text{o.\,t.,}  
\end{align}
where,
\begin{align}
    \label{value_of_a,b,c}
    a &= \cos^2\alpha ~ U\rho_{AS}U^\dagger, \nonumber \\
    b &= \cos\alpha\sin\alpha ~ U\rho_{AS}\widetilde{U}^\dagger,
    \nonumber \\
    c &= \sin^2\alpha ~ \widetilde{U}\rho_{AS}\widetilde{U}^\dagger,~
    \mbox{and}
    \nonumber \\
    \tau &= \frac{a+b+b^\dagger+c}{2}.
\end{align}
After postselecting the control qubit in state $\ket{+}_C$ we obtain,
\begin{gather}
     \rho^f_{AS} 
    = \frac{\tau}{\gamma},
    ~\mbox{where}~
    \nonumber \\
    \gamma = \Tr[\tau] =  \frac{1}{2}\left( 1 + \sin 2\alpha\text{Tr}\left[\widetilde{U}^\dagger U\rho_{AS}\right] \right),
    \label{prob-eqn}
\end{gather}
and the final system state is,
\begin{align}
\label{system-final-state}
    \rho_S^f &= \Tr_{A} (\rho^f_{AS}) = 
    \frac{\Tr_A[\tau]}{\gamma}.
\end{align}
Quantum channels must be trace-preserving and linear for them to be valid physical maps. 
From Eqs. \ref{prob-eqn} and \ref{system-final-state}, we have $\Tr[\rho_S^f]=1=\Tr[\rho_S]$, and therefore the channel $\varphi$ is trace-preserving.  
In the following, we assume $\rho_{A} = \ketbra{0_n}{0_n}_A$ and demand the linearity condition and arrive at constraints for the constituent unitaries.

\textit{Theorem:} If $\widetilde{U}^\dagger U$ is proportional to $\mathbb{I}$ in the subspace of $\rho_A = \ketbra{0_n}{0_n}_A$, then the superposed channel is linear.

\textit{Proof:} From Eq. \ref{prob-eqn}, we see that if $\widetilde{U}^\dagger U$ proportional to $\mathbb{I}$ in the subspace of $\rho_A = \ketbra{0_n}{0_n}_A$, then 
$\gamma$ will be independent of $\rho_S$ and hence $\varphi$ will be linear.

\textit{Corollary:} The above theorem implies that $\displaystyle \sum_{i=0}^{m-1} \widetilde{K}_i^{\dagger}K_i \propto \mathbb{I}$.

\textit{Proof:} $U$ and $\widetilde{U}^\dagger$ have the form,
\begin{equation}
    U = \left[ \begin{array}{c} K_0 \\ K_1 \\ \vdots \\ K_{m-1} \end{array} \middle| W \right],
    \quad
    \widetilde{U}^\dagger = \left[ \begin{array}{cccc} \widetilde{K}_0^\dagger & \widetilde{K}_1^\dagger & \cdots & \widetilde{K}_{m-1}^\dagger\\ \hline \multicolumn{4}{c}{V^\dagger} \end{array} \right],
\end{equation}
where $W$ and $V$ are rectangular matrices such that $U$ and $\widetilde{U}$ are unitary operators. 
The $jk^\mathrm{th}$ element of $\widetilde{U}^\dagger U$ for $j,k\in\{0,1\}$ is given by
\begin{equation}
    w_{jk} = (\widetilde{U}^\dagger U)_{jk} = \sum_{\mu=0}^1 \sum_{i=0}^{m-1}  (\widetilde{K}_i)_{\mu j}^* (K_i)_{\mu k}.
    \label{eq:w}
\end{equation}
The condition $_A\bra{0_n}\widetilde{U}^\dagger U\ket{0_n}_A \propto \mathbb{I}$ implies that, for $j,k\le 1$, $w_{jj}$ are all equal  and all $w_{jk}$ vanish for $j\neq k$, thereby leading to 
\begin{equation}
    \label{eqn-linearity-condition}
    \sum_{i=0}^{m-1}\widetilde{K}_i^\dagger K_i \propto \mathbb{I}.
\end{equation}
This is a sufficient condition for the superposed $\varphi$ channel to be linear.

\subsection{Superposition of General Pauli Channels}
\label{pauli-super-sect}
A general Pauli channel is defined by
\begin{equation}
    \label{gen-pauli-channel}
    \Lambda(\rho) = \sum_{i=0}^3~ p_i~ \sigma_i\rho\sigma_i
\end{equation}
where $\sigma_0 = \mathbb{I}$,  $\sigma_i$ for $i\in \set{1,2,3}$ are the Pauli-X, Y, Z matrices respectively, and error probabilities $p_i$ satisfy $\sum_{i=0}^3 p_i = 1$ to ensure the positivity of the map with corresponding Kraus operators $K_i = \sqrt{p_i}\sigma_i$.  Since there are four Kraus operators, we need  two ancilla qubits to realize the general Pauli channel.
The corresponding Stinespring dilation unitary  is
\begin{align}
    U(\bm{p}) &= \ketbra{00}{00}\otimes\sqrt{p_0}~\mathbb{I} + \ketbra{01}{00}\otimes\sqrt{p_1}~\sigma_1 
    \nonumber\\ 
    &+ \ketbra{10}{00}\otimes\sqrt{p_2}~\sigma_2 + \ketbra{11}{00}\otimes\sqrt{p_3}~\sigma_3 + \text{o.t},
    \label{eq-uforgenpauli}
\end{align}
where $\bm{p} = (p_0, p_1, p_2, p_3)$.

We now construct a  superposition of two Pauli channels with probabilities $\bm{p}$ and $\bm{\widetilde{p}}$ and corresponding Stinespring dilation unitaries $U(\bm{p})$ and $\widetilde{U}(\bm{{\widetilde{p}}})$. Such a superposed channel is a valid quantum channel since Eq. \ref{eqn-linearity-condition} is satisfied in general.
From Eq. \ref{value_of_a,b,c} we get
\begin{align}
    &\Tr_A[a] = \cos^2\alpha \sum_{i=0}^3 p_i ~\sigma_i \rho_S \sigma_i, \nonumber\\
    &\Tr_A[b] = \cos\alpha\sin\alpha\sum_{i=0}^3 \sqrt{p_i\widetilde{p}_i}~ \sigma_i \rho_S \sigma_i,~\mbox{and} \nonumber \\
    &\Tr_A[c] = \sin^2\alpha \sum_{i=0}^3 \widetilde{p}_i~ \sigma_i \rho_S \sigma_i.
\end{align}
Using Eqs. \ref{prob-eqn} and \ref{system-final-state} we obtain,
\begin{align}
 \label{gen-pauli-final-state}
  \gamma &= \frac{1}{2}\left(1+\sin{2\alpha}\sum_{i=0}^3 \sqrt{p_i\widetilde{p_i}}\right)
  ~\mbox{and}
  \nonumber \\
    \rho_S^f &= \frac{1}{2\gamma} 
    \sum_{i=0}^3\left(\cos{\alpha}~\sqrt{p_i} + \sin{\alpha}~\sqrt{\widetilde{p}_i}\right)^2 \sigma_i\rho_S \sigma_i.
\end{align}
Thus, the superposed channel of two Pauli channels is a valid, completely positive, and trace-preserving Pauli channel for all values of $\alpha$ with effective error probabilities
\begin{equation}
\mathbbm{p}_i = \frac{1}{2\gamma}(\sqrt{p_i}\cos{\alpha} + \sqrt{\widetilde{p_i}}\sin{\alpha})^2, 
\label{eq-pisup}
\end{equation}
satisfying $ \sum_{i=0}^{3} \mathbbm{p}_i = 1$.

Our goal is to measure the expectation values $\expv{\sigma_1}{}$, $\expv{\sigma_2}{}$, $\expv{\sigma_3}{}$ of the system state $\rho_S^f$ in Eq. \ref{system-final-state} after passing through the superposed channel.
As seen in Eq. \ref{system-final-state}, it involves two parts - measuring the unnormalized expectation values and normalizing them after measuring the factor $\gamma$.

To measure $\gamma$, we prepare the maximally mixed initial state $\rho_{SA} = \mathbb{I}/2^{n+1}$ and implement the quantum circuit shown in Fig. \ref{fig:circuits} (c).  
Using Eq. \ref{eq:w}, for a general Pauli channel acting on the Hilbert space $\mathcal{H}_A \otimes \mathcal{H}_S$, we see that $U$ and $\widetilde{U}$ are such that,

\begin{align}
w_{00} + 
w_{11} 
    &= \sum_{i=0}^3 \sum_{j=0}^1 (\tilde{K}_i)^*_{j0}~(K_i)_{j0} + \sum_{i=0}^3 \sum_{j=0}^1 (\tilde{K}_i)^*_{j1}~(K_i)_{j1} \nonumber \\
    &= 2\sum_{i=0}^3\sqrt{p_i \widetilde{p}_i}.
\end{align}
The normalization factor $\gamma$ can be determined by the conditional readout of the diagonal elements 
$\proj{0_2} \otimes \mathbb{I}/2 \otimes \sigma_3$ of the final density matrix $\rho^{f'}$ of circuit in Fig. \ref{fig:circuits} (c), since
\begin{align}
   \label{eqn-proof-add-coh-gives-gamma}
\Tr[\rho^{f'}(\proj{0_2} \otimes \mathbb{I}/2 \otimes \sigma_3)]    &=\frac{2\cdot\sin2\alpha(w_{00} + w_{11})}{8} \nonumber \\
    &= \frac{1}{2}\sin2\alpha\cdot\sum_{i=0}^3\sqrt{p_i\widetilde{p}_i} 
    \nonumber \\
    &= \gamma-\frac{1}{2}, ~\mbox{using Eq. \ref{gen-pauli-final-state}}.
\end{align}

In subsequent sections, we consider the specific cases of superposing two dephasing channels as well as superposing two depolarization channels.

\section{Superposition of dephasing channels}
\label{dephasing-super-sect}
A dephasing channel with strength $p \in [0,1]$ is a specific example of a general Pauli channel described in Eq. \ref{gen-pauli-channel} with $p_0 = 1-p,~ p_3 = p$, and $p_1 = p_2 = 0$. 
The Stinespring dilation unitary of Eq.~\ref{eq-uforgenpauli} with a single ancilla now takes the form
\begin{equation}
U(p) = 
    \begin{bmatrix}
        \sqrt{1-p} & 0 & \sqrt{p} & 0 \\
        0 & \sqrt{1-p} & 0 & \sqrt{p} \\
        \sqrt{p} & 0 & -\sqrt{1-p} & 0 \\
        0 & -\sqrt{p} & 0 & \sqrt{1-p}
    \end{bmatrix}.
    \label{eq:Uofp}
\end{equation}
In this section, we analyze the effects of superposing two dephasing channels of different strengths $p$ and $\widetilde{p}$. 
Using Eqs. \ref{gen-pauli-final-state} and  \ref{eq-pisup}, we find that the superposed channel is also a dephasing channel with effective strength 
\begin{align}
   \mathbbm{p} &= \dfrac{1}{2\gamma}(\sqrt{p}\cos{\alpha}+\sqrt{\widetilde{p}}\sin{\alpha})^2, ~\mbox{where} 
   \nonumber \\
   \gamma &= \frac{1}{2}\left(1+ \sin2\alpha~\left(\sqrt{p\widetilde{p}} + \sqrt{(1-p)(1-\widetilde{p})}\right)\right).
   \label{eq:effpandgamma}
\end{align}
In the following, we discuss the experimental demonstration of the above.

 \subsection{NMR realization of superposition of two dephasing channels}

 \begin{figure*}
    \centering
    \includegraphics[width=14cm, keepaspectratio]{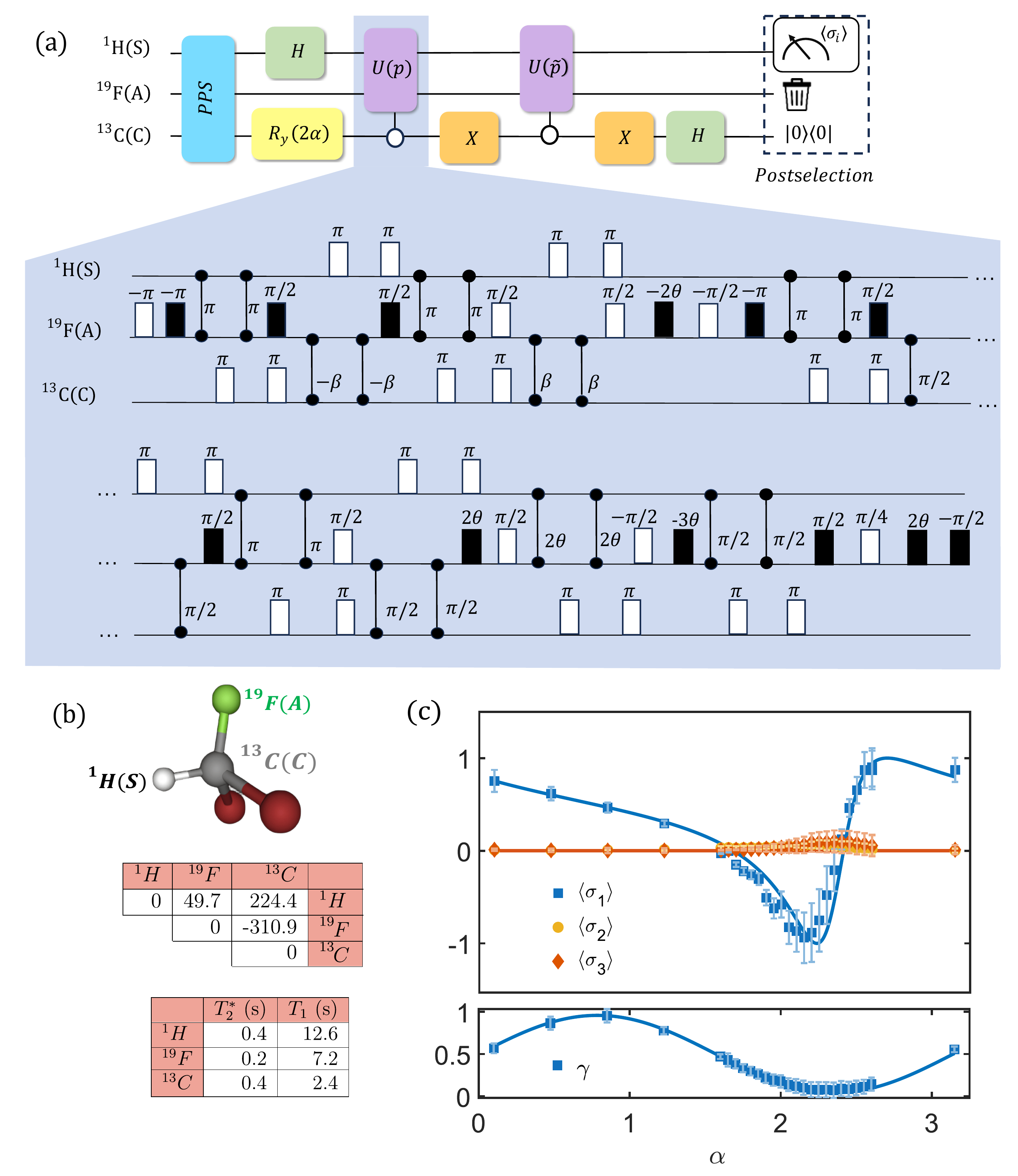}
    \caption{NMR realization of superposition of two dephasing channels.  (a) Pulse sequence implementing control unitaries $U(p)$ and $U(\widetilde{p})$ of the quantum circuit in Fig. \ref{fig:circuits} (b). Here,  $^1\mathrm{H}$, $^{19}\mathrm{F}$, and $^{13}\mathrm{C}$ are the system, ancilla, and control qubits respectively. White and black rectangles represent x- and y-pulses with pulse angles mentioned on top of them.
    Vertical lines joining qubits $a$ and $b$ represent angles $\Theta\in\{\pi,\pi/2,\pm\beta,2\theta\}$ realized by coupling evolution delays $\Theta/(4\pi J_{ab})$. 
    For negative $\Theta$, a delay of $3|\Theta|/(4\pi J_{ab})$ was used. 
    For the dephasing channel with strength $p$, $\beta = \tan^{-1}\sqrt{p/(1-p)}$, and $\theta = {\beta}/{2}$.  The normalization factor $\gamma$ is measured using a similar pulse sequence with different initial states as indicated in Fig. \ref{fig:circuits} (c).  (b) The molecular structure of DBFM.  The top table shows the values of $J$ coupling strengths for DBFM (top table, in Hz), and the bottom table lists $T_2^*$, $T_1$, relaxation time constants.  (c) The top box plots $\langle\sigma_i\rangle$ versus $\alpha$ for the superposition of two dephasing channels with strengths $p=0.1$ and $\widetilde{p} = 0.45$ acting on the initial state $\ketbra{+}{+}$ of the system qubit. The solid lines are numerical simulations, while the markers show experimental values with error bars. The bottom box plots $\gamma$ versus $\alpha$. The P1-error bars \cite{p1-error} of length one standard deviation were calculated considering RF inhomogeneity and noise level in the NMR spectra. Note that, for $\alpha = 1.68$ the two channels constructively interfere, thereby destroying the coherence, and for $\alpha=2.70$ they destructively interfere so as to preserve the coherence.}
    \label{dephasing-expt}
\end{figure*}

Here we utilize a 3-qubit register formed by three spin-$1/2$ nuclei 
$^1$H, $^{19}$F, and $^{13}$C, of 13C-dibromofluoromethane (DBFM) (see Fig. \ref{dephasing-expt} (b)) dissolved in acetone and maintained at an ambient temperature of 298 K inside an NMR magnet with a homogeneous magnetic field strength of 11.7 T. Under the high-temperature and  secular approximation, the  Hamiltonian in the triply resonant rotating frame is 
\begin{equation}
    H_\mathrm{DBFM} = \frac{\hbar\pi}{2} \left(J_\mathrm{HF} \sigma_3^\mathrm{H}\sigma_3^\mathrm{F}
    +
    J_\mathrm{FC} \sigma_3^\mathrm{F}\sigma_3^\mathrm{C}
    +
    J_\mathrm{HC} \sigma_3^\mathrm{H}\sigma_3^\mathrm{C}
    \right),   
\end{equation}
where constants $J$ are indirect spin-spin coupling strengths as tabulated in Fig.~\ref{dephasing-expt} (b). 
The entire experimental procedure is summarized in Fig.~\ref{dephasing-expt} (a).
Here we initialize the system in $\ket{000}$ pseudopure state  using the three-qubit spatial averaging sequence \cite{dbfm-pulse-seq, pps-prep} before implementing the quantum circuit of Fig.~\ref{fig:circuits} (b). To simplify the NMR pulse sequence, we implemented the phase-shifted Stinespring dilation unitary
\begin{equation}
U(p) = 
    \begin{bmatrix}
        -i\sqrt{1-p} & 0 & -i\sqrt{p} & 0 \\
        0 & \sqrt{1-p} & 0 & \sqrt{p} \\
        -i\sqrt{p} & 0 & i\sqrt{1-p} & 0 \\
        0 & -\sqrt{p} & 0 & \sqrt{1-p}
    \end{bmatrix}
\end{equation}
as shown in Fig.~\ref{dephasing-expt} (a).
The modified unitary has the same effect as that of Eq. \ref{eq:Uofp}, but with interchanged observables, i.e., $\sigma_1 \rightarrow \sigma_2$ and $\sigma_2 \rightarrow \sigma_1$.

We implement the superposition of two dephasing channels of strengths $p=0.1$ and $\widetilde{p}=0.45$ for a range of $\alpha$ values, and each time we monitor the effect of the superposed channel on the $\ket{+}$ initial state of the system qubit. 
In the 
top box of Fig.~\ref{dephasing-expt} (c),
the Pauli expectation values  $\langle \sigma_1\rangle, \langle \sigma_2\rangle, \langle \sigma_3\rangle$ of the system qubit  are plotted against the parameter $\alpha$ that controls the strength of superposition.

We observe that the expectation value  $\langle \sigma_3 \rangle$ does not change with $\alpha$, which is typically expected from a dephasing channel. Interestingly, at $\alpha = 2.70\pm n\pi$, for integer $n$, the two noisy channels destructively interfere with each other, allowing the recovery of the original un-dephased state $\ket{+}$. At  $\alpha = 2.23 \pm n\pi$, the superposed channel outputs the orthogonal state $\ket{-}$. Furthermore, at $\alpha = 1.68 \pm n\pi$, the two dephasing channels constructively interfere to output the maximally mixed state through the superposed channel. These experiments  demonstrate the interference effect of superposing two dephasing channels.

The bottom box of Fig~\ref{dephasing-expt} (c) plots the normalization factor $\gamma$ versus $\alpha$.  Note that $\gamma$ is also the probability of postselection.  While $\alpha=2.7\pm n\pi$ cancels the two dephasing channels, we notice from Fig~\ref{dephasing-expt} (c) that the corresponding postsection probability $\gamma \sim 0.06$ is quite small. 
Thus, we get error-free qubits at the cost of a reduced ensemble size.  

In the following we discuss the experimentally superposing two depolarizing channels.

\section{Superposition of Depolarizing channels}
\label{depol-super-sect} 
 A depolarizing channel with strength $p$  
corresponds to 
\begin{equation}
\bm{p} = (\sqrt{1-p}, \sqrt{p/3}, \sqrt{p/3}, \sqrt{p/3})
\label{eq:depol}
\end{equation}
in Eqs. \ref{gen-pauli-channel} and \ref{eq-uforgenpauli}, and causes the Bloch sphere to contract towards the maximally mixed state as we increase the strength $p$. We now consider the superposition of two depolarizing channels with strengths $p$ and $\widetilde{p}$. From Eq. \ref{gen-pauli-final-state}, we find that the final state of the system after going through the superposed channel is

\begin{align}
    \nonumber
    \rho_S^f = \frac{1}{2\gamma} &\left[  (\sqrt{1-p}\cos{\alpha} + \sqrt{1-\widetilde{p}}\sin{\alpha})^2\rho_S + 
    \right. 
    \nonumber \\
    &\left. \left(\sqrt{\frac{p}{3}}\cos{\alpha} + \sqrt{\frac{\widetilde{p}}{3}}\sin{\alpha}\right)^2\sum_{i=1}^3\sigma_i\rho_S\sigma_i
    \right]. 
\end{align}

It is easy to verify that this superposed channel is also a depolarizing channel with strength,
\begin{align}
    \label{depol_super_strength}
    \mathbbm{p} &= \frac{3}{2\gamma}\left(\sqrt{\frac{p}{3}}\cos{\alpha} + \sqrt{\frac{\widetilde{p}}{3}}\sin{\alpha}\right)^2, 
\end{align}
where $\gamma$ happens to be the same as in Eq. \ref{eq:effpandgamma}.

\subsection{Superposing Entanglement Breaking (EB) Channels}
\label{ebc-super-sect}
A channel $\mathcal{E}$ is entanglement breaking (EB) if it acts on a subsystem of a bipartite state $\rho$ such that $(\mathcal{E}\otimes\mathbb{I})(\rho)$ is always separable for all $\rho$ \cite{EBC, qubit-EBC}. 
EB channels are of significant importance, as these channels can be simulated using classical channels and are equivalent to measure-and-prepare channels \cite{Holevo1998-ebc, qubit-EBC}, which only require local operations and classical communication (LOCC) \cite{qubit-EBC}.
Since entanglement is a central resource for quantum technologies, unintentional EB channels pose challenges for quantum information processing as well as quantum communication tasks like teleportation and quantum key distribution.
It was shown that if a channel is an EB channel, then its corresponding Choi state \cite{CHOI1975285, JAMIOLKOWSKI1972275} is separable \cite{EBC}. While amplitude damping and phase damping are EB channels only in the asymptotic limits of channel strengths, the depolarizing channel is an EB channel over a wide range of depolarizing strengths.
The Choi matrix for the depolarizing channel of Eq. \ref{eq:depol}
is given by
\begin{equation}
    C(p) = \begin{bmatrix}
        1-\frac{2}{3}p & 0 & 0 & 1-\frac{4}{3}p \\
        0 & \frac{2}{3}p & 0 & 0 \\
        0 & 0 & \frac{2}{3}p & 0 \\
        1-\frac{4}{3}p & 0 & 0 & 1-\frac{2}{3}p
    \end{bmatrix},
\end{equation}
which is isomorphic to a 2-qubit state, and hence the positivity of its partial transpose is the necessary and sufficient condition for separability \cite{separability_cond}. When $p \in [1/2, 1]$, the Choi state corresponding to $C(p)$ is separable, and hence in this regime, the depolarizing channel is an EB channel.

\begin{figure}
    \centering
    \includegraphics[width=8.5cm, keepaspectratio]{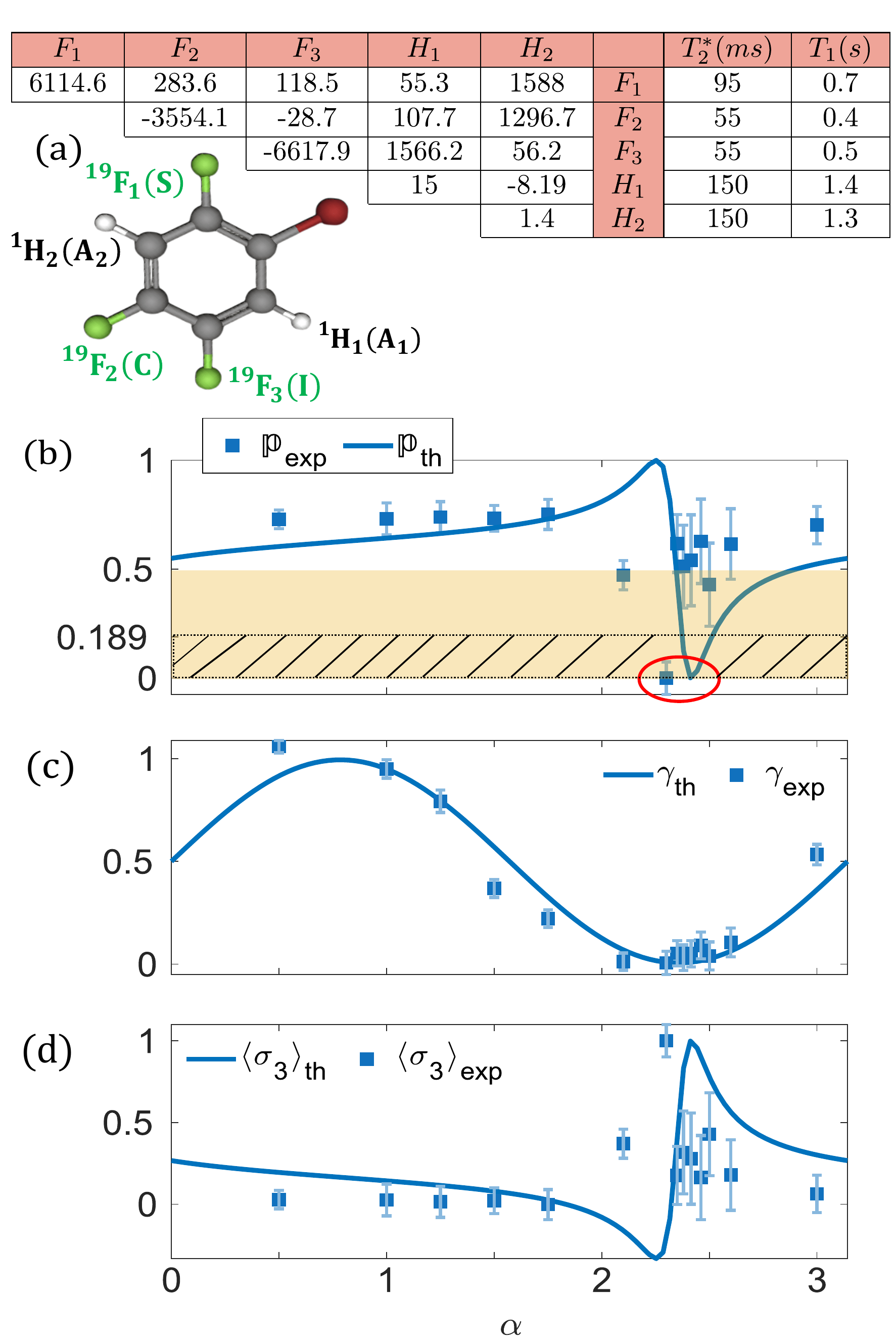}
    \caption{Experimental realization of superposition of two depolarizing channels of strengths $p=0.55$ and $\widetilde{p} =0.7$.
    (a) Molecular structure of BTFBz and table of resonance offsets $\nu_i$ (diagonal elements in Hz), effective coupling constants $J_{ij}+2D_{ij}$ (off-diagonal elements in Hz), and relaxation time constants. (b-d) The effective strength $\mathbbm{p}$ of superposed channel, the normalization factor $\gamma$, and expectation value $\langle \sigma_3 \rangle$, plotted versus the superposition parameter $\alpha$.  The lines and markers represent theory and experimental data. In all the cases, P1-error bars \cite{p1-error} of length one standard deviation were calculated considering RF inhomogeneity and thermal noise. In (b), the  shaded area represents the non-EB channel regime, and the hashed area represents the superactivation regime of the superposed channel.
    }
    \label{fig:ebc-expt}
\end{figure}

We now consider  superposing two EB channels and ask if the superposed channel can be a non-EB channel.  
The solid line in Fig.~\ref{fig:ebc-expt} (b) plots the effective theoretical strength $\mathbbm{p}$ 
(see Eq.~\ref{depol_super_strength}) of the channel obtained by superposing two depolarizing EB channels with strengths, $p= 0.55$ and $\widetilde{p} = 0.7$ versus the superposition parameter $\alpha$.
As discussed earlier, the superposed channel is also a depolarizing channel.  However, as seen from Fig.~\ref{fig:ebc-expt} (b), for $\alpha \in [2.35, 2.90] \pm n\pi$, the superposed channel is a non-EB channel as indicated by the shaded region in Fig.~\ref{fig:ebc-expt} (b).  Thus, it is interesting to note that the superposition of two EB channels can be a non-EB channel for certain values of $\alpha$.

We now analyze the superposed channel in terms of its quantum capacity, which refers to the highest rate at which quantum information can be reliably transmitted \cite{Mark_Wilde_Book}. An EB channel outputs a separable state for any input entangled state, and hence it is a zero-capacity channel \cite{qswitch-comm-2}. 
Multiple works studied the lower and upper bounds of quantum capacity of a depolarizing channel \cite{q-capacity-depol1, q-capacity-depol2, q-capacity-depol3, q-capacity-depol4}. It is shown that when $p \le 0.189$, the channel will have a positive lower bound of quantum capacity \cite{q-capacity-depol2}.
Superactivation of quantum capacity  refers to the phenomenon when combining two zero-capacity channels results in a positive capacity channel \cite{superactivation1}.  
As indicated by the hashed region  in Fig.~\ref{fig:ebc-expt} (b), the effective channel strength $\mathbbm{p} \le 0.189$ for $\alpha \in [2.37,2.51]\pm n\pi$, corresponding to a positive quantum capacity, thereby establishing superactivation.
Specifically at $\alpha = 2.416\pm n\pi$ (highlighted by the ellipse), the superposition of two EB channels leads to vanishing effective channel strength $\mathbbm{p}=0$,  corresponding to an ideal channel.  

Thus, two zero-capacity channels can destructively interfere and result in the \textit{perfect activation} of quantum capacity \cite{qswitch-comm-2}.  
In other words,
two individual channels unsuitable for transmitting quantum information, together in superposition can enable perfect communication.  However, like in the previous case of superposing dephasing channels, we see a trade-off between the ensemble size and the transmission quality of the superposed channel. Here we find that the probability $\gamma$ is as low as 0.006 at $\alpha = 2.356 \pm n\pi$ within the superactivation region.

While Ref. \cite{qswitch-comm-2}  conceptualized the perfect activation of quantum capacity   with a quantum switch, the same with superposing two independent noise channels with separate ancillary registers was argued to be infeasible. 
Nonetheless, here we showed that if the superposition of two noisy channels is performed with a common ancillary register, we can achieve perfect activation of the quantum capacity.

\subsection{NMR realization of superposition of two depolarizing channels}
We now utilize a 5-qubit register consisting of three $^{19}$F and two $^1$H nuclear spins of 1-bromo-2,4,5-trifluorobenzene (BTFBz, Fig.~\ref{fig:ebc-expt} (a)) partially oriented in the nematic liquid crystal N-(4-methoxybenzylidene)-4-butylaniline (MBBA). The sample was maintained at an ambient temperature of 298 K inside an NMR magnet with a homogeneous magnetic field of 11.7 T.  
Under the high-temperature and high field secular approximation, the  Hamiltonian in the triply resonant rotating frame is given by
\begin{equation}
    H_\mathrm{BTFBz} = -\hbar\pi\sum_{i}\nu_{i}{\sigma_3^i} + \frac{\hbar\pi}{4}\sum_{i\neq j} (J_{ij}+2D_{ij}) \sigma_3^i\sigma_3^j,   
\end{equation}
where $i,j\in[\mbox{F}_1,\mbox{F}_2,\mbox{F}_3,\mbox{H}_1,\mbox{H}_2]$, and
$\nu_i$, $J_{ij}$, $D_{ij}$ are the resonance offsets, indirect spin-spin coupling constants, and the partially averaged dipolar coupling constants as tabulated in Fig.~\ref{fig:ebc-expt} (a).
We first prepare a pair of pseudo-pure states (POPS) \cite{pops-method} 
    $\rho_0 = \ketbra{0000}{0000} \otimes \sigma_{3}$
using $\mbox{F}_3$ as the labeling qubit. 
We consider $\mbox{F}_1$ is as the system ($S$) qubit, $\mbox{F}_2$ as the control ($C$) qubit and  $\mbox{H}_1$, $\mbox{H}_2$ as the ancillas ($A_1$, $A_2$) respectively.  
We now implement the quantum circuits of Fig. \ref{fig:circuits} (b) and (c) to realize the superposition of two depolarising channels having strengths $p = 0.55$ and $\widetilde{p} = 0.7$ and to determine $\gamma$.
We apply a $\alpha$ rotation $R_y(2\alpha)$ on $C$ to prepare it into a superposed state $\cos\alpha\ket{0} + \sin\alpha\ket{1}$ and apply the controlled unitary
\begin{equation}
    U(p) \otimes\ketbra{0}{0}_C \otimes \mathbb{I}_I~ +~ U(\widetilde{p}) \otimes\ketbra{1}{1}_C \otimes \mathbb{I}_I.
\end{equation}
where $U(p)$ and $U(\widetilde{p})$  are the Stinespring dilation unitaries.  For the depolarizing channel acting on the joint space $\mathcal{H}_{A_1}\otimes \mathcal{H}_{A_2} \otimes \mathcal{H}_S$ 
\begin{equation}
\label{eqn-depol-stinespring-U}
    U(p) = 
\begin{bmatrix}
    q_0\mathbb{I} & q\mathbb{I} & q\mathbb{I} & q\mathbb{I}\\
    q\sigma_1            & (1-f)\sigma_1      & -f\sigma_1         & -f\sigma_1         \\
    q\sigma_2            &  -f\sigma_2        & (1-f)\sigma_2      & -f\sigma_2         \\
    q\sigma_3            &  -f\sigma_3        & -f\sigma_3         & (1-f)\sigma_3  
\end{bmatrix},
\end{equation}
where $q_0 = \sqrt{1-p}$, $q = \sqrt{{p}/{3}}$, and $f = {q^2}/{(1-q_0)}$. Subsequently, a Hadamard gate is applied on $C$ before postselecting it in the $\ket{0}$ state. We realized all these operations using amplitude- and phase-modulated RF pulses designed with the GRAPE algorithm \cite{grape_simpson, grape1}. 
The normalization factor $\gamma$, and normalized expectation values $\langle \sigma_3 \rangle$ of the system qubit are plotted versus $\alpha$ in Figs. \ref{fig:ebc-expt} (c) and (d). After dividing by $\gamma$, the experimental $\langle \sigma_3 \rangle$ values are further normalized by their highest value. 
Although experimental $\gamma$ values appear to follow the theoretical curve, experimental $\langle \sigma_3 \rangle$ values show significant deviations from theory, especially for $\gamma$ values close to zero.

Under depolarizing channel, the Bloch sphere shrinks uniformly, and accordingly each expectation value
$\langle \sigma_i \rangle$ reduces by the factor $(1-4p/3)$ \cite{Nielsen_Chuang_2010}.
We experimentally measure $\langle \sigma_3\rangle_{\rho_0}$ and $\langle \sigma_3\rangle_{\varphi(\rho)}$ respectively for before and after the system state has gone through the superposed channel and thereby  retrieve the effective strength of the superposed channel
\begin{equation}
    \label{p_and_exp_depol}
    \mathbbm{p} = \frac{3}{4}\cdot\left(1-\frac{\langle \sigma_3\rangle_{\varphi(\rho)}}{\langle \sigma_3\rangle_{\rho_0}}\right).
\end{equation}
The experimentally obtained values of $\mathbbm{p}$ are plotted versus $\alpha$ in Fig. \ref{fig:ebc-expt} (b).
Although we see significant deviations of $\mathbbm{p}$ values from theory, it is interesting to see a couple of points showing non-EB superposed channel, and particularly one point corresponding to the perfect activation within the errorbar as indicated by the ellipse in Fig. \ref{fig:ebc-expt} (b).  

\section{Conclusions}
\label{conclusions}
In this work, we extended the framework of superposition of unitaries to arrive at an alternative structure for superposing quantum channels. We utilized a single control qubit to superpose two Stinespring dilation unitaries acting on a common ancillary register to realize a valid superposed quantum channel.
The question arises if all channels can be superposed in such a fashion. We arrived at a set of conditions that need to be satisfied for the superposed channel to be a valid quantum channel. We then considered the superposition of two general Pauli channels and took specific examples of dephasing and depolarizing channels. We showed the interference of channels and the recovery of the original state through the superposition of dephasing channels. Moreover, we showed the superactivation and perfect activation of quantum capacity by superposing two zero-capacity channels in the form of entanglement-breaking depolarization channels. 

We demonstrated these theoretical results experimentally using NMR. Using a three-qubit NMR register, we demonstrated the superposition of two dephasing channels and observed both constructive and destructive interferences of dephasing channels. Using a five-qubit NMR register, we demonstrated the superposition of two entanglement-breaking depolarization channels and observed an effective non-entanglement-breaking superposed channel. Within  experimental error bars, we also observed superactivation as well as perfect activation.

Our work demonstrates a novel approach to coherent control of quantum channels, which can lead to novel applications in quantum information processing and quantum communication.  It is interesting to explore other variants such as superposition of two different channels, superposition of three or more channels, and iterative applications of superposed channels. 

\acknowledgments
Authors acknowledge valuable discussions with Prof. Usha Devi and Dr. Karthik. D. B. acknowledges the INSPIRE fellowship from the Department of Science and Technology (DST), Govt. of India. V. V. acknowledges the University Grants Commission (UGC)
fellowship MR22031373. We thank the National Mission
on Interdisciplinary Cyber-Physical Systems for funding
from the DST, Government of India, through the I-HUB
Quantum Technology Foundation, IISER-Pune.

\bibliography{bibliography}
\end{document}